\documentclass[12pt,preprint]{aastex}

\shorttitle{Plasma instabilities as a result of charge exchange in the shock downstream}

\shortauthors{Ohira et al}

\begin{document}

\title{Plasma instabilities as a result of charge exchange in the downstream region of supernova remnant shocks}

\author{Yutaka Ohira\altaffilmark{1}, Toshio Terasawa\altaffilmark{2} and Fumio Takahara\altaffilmark{1}}

\begin{abstract}
H$\alpha$ emission from supernova remnants (SNRs) implies the existence of neutral hydrogen in the circumstellar medium. Some of the neutral particles penetrating the shock are ionized by the charge exchange process and make a cold ion beam in the shock downstream region. We perform linear analyses of collisionless plasma instabilities between the cold beam and the hot downstream plasma. We find that, under typical SNR conditions, either the resonant instability or the Weibel instability is the most unstable. This mechanism may amplify the magnetic field to more than 100${\rm \mu G}$ and changes the shock structure. As a result, the radio spectrum and the large magnetic field can be explained, apart from the widely discussed Bell's mechanism.
\end{abstract}

\keywords{magnetic fields; plasmas; shock waves; supernova remnants}

\altaffiltext{1}{Department of Earth and Space Science, Graduate School of Science, Osaka University, 1-1 Machikaneyama-cho, Toyonaka, Osaka 560-0043, Japan; yutaka@vega.ess.sci.osaka-u.ac.jp}
\altaffiltext{2}{Interactive Research Center for Science and Department of Physics, Tokyo Institute of Technology, Tokyo 152-8551, Japan}

\section{Introduction}
Supernova remnants (SNRs) are observed in various bands from radio to gamma-ray. 
X-ray observations provide the evidence that electrons are accelerated to highly relativistic energies in SNR shocks \citep{koy95}. Moreover, it is believed that ions are also accelerated to the "knee" ($\sim10^{15}$ eV) of the Galactic cosmic ray spectrum as the origin of galactic cosmic rays. X-ray observations of SNRs also suggest the magnetic field amplification up to $10^{2-3}{\rm \mu G}$ in the vicinity of the collisionless shock front \citep{vin03, ber03, bam05, uch07}. Theoretically, several mechanisms for magnetic field amplification have been proposed, one of which is the cosmic ray driven instability, whereby cosmic rays diffuse to the upstream region where the resonant and the nonresonant instabilities become unstable and cause fluctuations in the magnetic field \citep{luc00, bel04, rev06, rev07}. Recently, the nonresonant instability has been investigated with Particle In Cell simulations \citep{nie08,riq08,ohi09} and MHD simulations \citep{bel04, bel05, rev08, zir08}, but \citet{ohi09} claim that the periodic boundary simulations cannot describe nonlinear features. Thus, the amplitude and the wave spectrum of the magnetic fields excited by the nonresonant instability in the downstream region are open issues. The other magnetic field amplification mechanism is fluid type instabilities in the downstream region \citep{gia07, ino09}. These instabilities generate waves with longer wavelength than plasma kinetic instabilities.

In this letter, we propose yet another mechanism of the magnetic field amplification, where neutral particles in the partially-ionized interstellar matter (ISM) play a significant role. The existence of neutral particles has been identified in many young SNRs from the detection of H$\alpha$ emissions. Since SNR shocks are collisionless, neutral particles can penetrate the shock front freely without deceleration. As a result, a significant relative velocity exists between the plasma particles and neutral particles in the shock downstream region, which is given by $v_{\rm rel}$=$3v_{\rm sh}/4$ ($v_{\rm sh}$ is the shock velocity). Through the charge exchange process in the downstream region, shocked hot protons and cold neutral hydrogen atoms are converted to hot neutral hydrogen atoms and cold protons, respectively \citep{hen07, van08}. It is noted that the H$\alpha$ emissions from SNR consist of two components with narrow and wide line widths \citep{che78, che80, smi91}. The narrow-width component originates from cold hydrogen atoms (before charge exhange) excited by heated proton and electron collisions, while the latter wide-width component originates from hot hydrogen atoms produced by the charge exchange process.

The neutral fraction of the ISM around supernovae is often found to be order of unity. For example, \citet{gha00, gha02} have shown that the neutral fractions are about 0.8 and 0.1 in Tycho and SN1006 environments. Therefore, the charge-exchanged cold protons can have energy density comparable to that of the hot shocked protons, so that their effects on the downstream physical condition should be quite important. Such a condition is similar to interacting region between the solar wind and a comet or interstellar neutral particles where several collisionless plasma instabilities have been considered \citep{wu72}.

We first provide the background plasma condition (\S 2). We then perform linear analyses of the collisionless plasma instabilities for a parallel shock (\S 3) and discuss impacts on the magnetic field amplification, the shock structure and the particle acceleration (\S 4). Studies of the perpendicular or oblique shocks will be presented in another paper.

\section{Background plasma condition in the shock downstream region}
Although collisionless shock structures have not been understood completely, the dissipation scale is thought to be about the ion gyroradius, $r_{\rm g, i}\sim 10^{8-10}$cm, which is much shorter than the ionization scale, $l_{\rm ion}\sim 10^{15-17}$cm \citep{van08}. Therefore the charge exchange process occurs after the upstream plasma density is compressed according to the Rankin-Hugoniot relation. Interestingly, this ionization scale is comparable to the spatial scale of X-ray filaments \citep{bam05}. In this letter, we treat only the parallel shock, so that the magnetic field of ISM $B_{\rm ISM}$ is not compressed. For the particle components in the downstream region, we consider neutral hydrogen atoms (density $n_{\rm H}$), shocked hot protons ($n_{\rm p, hot}$) and electrons ($n_{\rm e}$), and charge-exchanged cold protons ($n_{\rm p, cold}$).
We choose typical values for young SNRs as,
\begin{equation}
n_{\rm p,hot}=n_{\rm e} = 1 {\rm cm}^{-3}, \ \ B_{\rm ISM}=3{\rm \mu G}, 
\end{equation}
and set $v_{\rm sh}=0.01c$. The density of cold protons $n_{\rm p, cold}$ becomes higher as the neutral hydrogens flow to the downstream, and is a function of the elapsed time after passing over the shock front,
\begin{equation}
\frac{D n_{\rm p,cold}}{Dt} =n_{\rm H} n_{\rm p,hot}\sigma_{\rm C.E.}v_{\rm rel},
\end{equation}
where $\sigma_{\rm C.E.}$ is the cross section of charge exchange between the hydrogen atom and shocked proton. We set $\sigma_{\rm C.E.}=10^{-15}{\rm cm}^2$ as a typical value,
so that the charge exchange time $(n_{\rm p,hot}\sigma_{\rm C.E.}v_{\rm rel})^{-1}=4.4\times 10^6 {\rm sec}$. From equation (2), we have
\begin{equation}
n_{\rm p,cold} = n_{\rm H0} \left(\frac{t}{4.4\times 10^6 {\rm sec}}\right),
\end{equation}
where we set $n_{\rm H}$ to be constant at $n_{\rm H0}$, the atomic hydrogen density in the ISM. This approximation holds as long as $n_{\rm p,cold}\ll n_{\rm H0}$. To satisfy the current neutrality, in the downstream electron plasma rest frame, drift velocities of two component of protons are 
\begin{eqnarray}
v_{\rm d,hot}&=& -\frac{3n_{\rm p,cold}}{4(n_{\rm p,hot}+n_{\rm p,cold})}v_{\rm sh} \nonumber \\ 
v_{\rm d,cold}&=&\frac{3n_{\rm p,hot}}{4(n_{\rm p,hot}+n_{\rm p,cold})}v_{\rm sh}.
\end{eqnarray}
The temperatures are
\begin{equation}
T_{\rm p,hot}= \frac{3}{16}m_{\rm p}v_{\rm sh}^2=18.75{\rm keV}, \ \ T_{\rm p,cold}=1{\rm eV}, \ \ {\rm and} \ \ T_{\rm e} =0.03T_{\rm p,hot},
\end{equation}
where we assume the downstream electron and proton temperature ratio $T_{\rm e}/T_{\rm p,hot}=0.03$ as the typical value of young SNRs \citep{gha07, van08, ohi08}.

\section{Linear analysis of collsionless plasma}
In this section, we perform linear analysis by treating $n_{\rm p,cold}$ as a free parameter. We consider only the cases where the wave vector ${\bf k}$ is purely parallel or perpendicular to the magnetic field. To obtain the maximum growth rate of the instabilities as a function of $n_{\rm p,cold}/n_{\rm p,hot}$, we solve numerically the standard collsionless plasma dispersion relation with the drift velocity and the thermal velocity \citep{ich73} because the thermal effect is important in the downstream region.

\subsection{${\bf k} \parallel {\bf B}$}
For parallel propagation, there are three unstable modes. One is the ion acoustic instability which is an electrostatic mode. The dispersion relation is given by
\begin{eqnarray}
&&0=1+\sum_s \frac{2\omega_{{\rm p},s}^2}{k^2v_{{\rm th},s}^2}
\left[1+\xi_s Z(\xi_s)\right], \nonumber \\
&&Z(\xi_s) = \frac{1}{\sqrt{\pi}}
\int_{-\infty}^{\infty}\frac{e^{-z^2}}{z-\xi_s}dz, \\
&&\xi_s = \frac{\omega-kv_{{\rm d},s}}{kv_{{\rm th},s}}, \nonumber
\end{eqnarray}
where the subscript $s$ represents particle species, here electron, hot proton, and cold proton, and $\omega_{{\rm p},s}, v_{{\rm d},s}$ and $v_{{\rm th},s}$ are the corresponding plasma frequency, drift velocity and thermal velocity, respectively, and $Z(\xi_s)$ is the plasma dispersion function. 

The others are the resonant and the nonresonant instabilies which are electromagnetic modes and can amplify the fluctuating magnetic field \citep{win84}. The dispersion relation for right handed electromagnetic waves is given by
\begin{eqnarray}
&&0=1-\left(\frac{ck}{\omega}\right)^2+\sum_s \left(\frac{\omega_{{\rm p},s}}{\omega}\right)^2
\frac{\omega -kv_{{\rm d},s}}{kv_{{\rm th},s}} Z(\xi_s) \nonumber \nonumber \\
&&\xi_{s}=\frac{\omega -kv_{{\rm d},s}+\Omega_{{\rm c},s}}{kv_{{\rm th},s}} ,
\end{eqnarray}
where $\Omega_{{\rm c},s}=q_s B/m_s c$ is the corresponding plasma cyclotron frequency ($q_{\rm p}=-q_{\rm e}=e$). The resonant instability satisfies the condition $Re[\omega] -kv_{\rm d,cold}+\Omega_{\rm c,p}=0$, so that the propagating direction is the same as the drift velocity of cold protons. On the other hand, the nonresonant instability does not satisfy the resonance condition and the propagating direction is opposite to that of the resonant mode.

\subsection{${\bf k} \perp {\bf B}$}
For perpendicular propagation, the Weibel instability with the uniform magnetic field is excited. In this condition, the unstable mode is the ordinary mode, that is, the fluctuating electric field is parallel to the magnetic field, so that the dispersion relation is given by
\begin{eqnarray}
&&0=1-\left(\frac{ck}{\omega}\right)^2-\sum_s \left(\frac{\omega_{{\rm p},s}}{\omega}\right)^2 \nonumber \\
&& \ \ \ \ -2\sum_s \left(\frac{\omega_{{\rm p},s}}{\omega}\right)^2\left[1+2\left(\frac{v_{{\rm d},s}}{v_{{\rm th},s}}\right)^2\right]\sum_{n=1}^{\infty}I_n(\lambda_s)e^{-\lambda_s}\frac{n^2\Omega_{{\rm c},s}^2}{\omega^2-n^2\Omega_{{\rm c},s}^2} \\
&&\lambda_{s}=\frac{kv_{{\rm th},s}}{\Omega_{{\rm c},s}} \nonumber,
\end{eqnarray}
where $I_n$ is the modified Bessel function of the first kind with order $n$.

\subsection{results}
Substituting  equations (1), (4) and (5) to the dispersion relations (6), (7) and (8), we obtain the growth rates of these instabilities. Figure 1 shows the maximum growth rates of these instabilities as a function of $n_{\rm p,cold}/n_{\rm p,hot}$, where the growth rates are normalized by $\Omega_{\rm c,p}$. Note that even if $n_{\rm p,cold}/n_{\rm p,hot}$ is large, the growth rate of the nonresonant instability is smaller than that of the resonant instability because the thermal velocity of the downstream plasma is comparable to the relative velocity $3v_{\rm sh}/4$. The Weibel instability is stabilized when its growth rate becomes smaller than the proton cyclotron frequency because the plasma motion is controlled by the uniform magnetic field. In our plasma condition, the Weibel instability is stabilized when $n_{\rm p,cold}/n_{\rm p,hot}$ is smaller than about $3\times 10^{-3}$. This critical density depends on thermal velocities, the uniform magnetic field and the drift velocities.

To find which instability becomes nonlinear first after the shock passage, we compare the growth rate of the instabilities with the growth rate of the cold proton density $t^{-1}$.  From equation (3), we plot $1/(t\Omega_{\rm c,p})$ as a function of $n_{\rm p,cold}/n_{\rm p,hot}$ in figure 1. The black solid and dashed lines show the case of $n_{\rm H0}=1{\rm cm}^{-3}$ and $20{\rm cm}^{-3}$, respectively. The black solid line and the growth rate of the resonant instability cross at $n_{\rm p,cold}/n_{\rm p,hot}\simeq 5\times10^{-4}$, so that first of all the resonant instability becomes nonlinear after passing over the shock front at around 50 times the proton cyclotron time. If the number density of hydrogen atoms in ISM $n_{\rm H0}$ is $\sim 20{\rm cm}^{-3}$ as observed in high density clumps in ISM \citep{hei03, ino08}, it is not the resonant instability but the Weibel instability that first reaches the nonlinear phase. Even when $n_{\rm H0}$ is as high as $20{\rm cm}^{-3}$, the collision frequency with neutral particles is about $10^{-7}{\rm sec}^{-1} \simeq 10^{-6}\Omega_{\rm c,p}$ \citep{rev07} which is much smaller than the growth rates, so that the collisional damping by the neutral particles is not important. Depending on the ISM density, either of the resonant or the Weibel instability works to amplify the magnetic field. If this is the case, the high magnetic field intensity found in some SNRs is explicable. Although figure 1 shows that the ion acoustic instability is the most unstable mode when $n_{\rm p,cold}/n_{\rm p,hot}>10^{-2}$, in fact it is not because cold protons have already been slightly heated by the waves excited by the resonant or the Weibel instabillity before $n_{\rm p,cold}/n_{\rm p,hot}$ attains such a value. For example, when $n_{\rm p,cold}/n_{\rm p,hot}$ becomes 0.1, if the temperature of cold protons becomes larger than about 50 times the initial temperature, the growth rate of the Weibel instability becomes larger than that of the ion acoustic instability. Even during such a large temperature increase of the cold protons, the kinetic energy of neutral atoms is almost conserved, so that its further conversion to the magnetic field energy remains possible.

\section{Summary and Discussion}
We performed linear analyses of collisionless plasma instabilities caused by cold proton beam generated by charge exchange process. We show that the resonant instability or the Weibel instability becomes nonlinear after passing over the shock front and the cold beam carries a large kinetic energy. For the estimation of the magnetic field amplified by the instabilities, we can use a simple order-of-magnitude estimation based on the simulations by \cite{kat08},
\begin{equation}
\delta B \simeq 100 \left(\frac{\varepsilon_{B}}{0.01}\right)^{1/2}\left(\frac{n_{\rm H0}}{0.4{\rm cm}^{-3}}\right)^{1/2}\left(\frac{v_{\rm sh}/c}{0.01}\right){\rm \mu G},
\end{equation}
where $\varepsilon_{B}$ is energy conversion parameter from the kinetic energy to the magnetic energy. Equation (9) is close to the suggestion by the X-ray observations. To make further discussion, however, we need more realistic simulations in which charge exchange process is properly taken into account.

Next, we discuss the modification of shock structure by the existence of neutral atoms. Neutral atoms of ISM can penetrate the shock front without the dissipation and the deceleration. Although we do not know how the mixing occurs after ionization, we speculate that eventually cold ions and hot ions have the same drift velocity which is one of the center of mass and that the relaxation scale is the ionizing scale. In this case, the far downstream velocity in the shock rest frame $v_{\rm down}$ is 
\begin{equation}
v_{\rm down} = \frac{1}{1+3\chi}v_{\rm sh},
\end{equation}
where $\chi$ is the ionization fraction of ISM. Hence, the spectral index of radio synchrotron from accelerated electrons $\alpha$ is
\begin{equation}
\alpha = \frac{1}{2\chi},
\end{equation}
where the acceleration mechanism is the diffusive shock acceleration so that $\alpha$ depends on the velocity ratio $v_{\rm sh}/v_{\rm down}$ \citep{kry77, axf77, bel78, bla78}. If $\chi=5/6$, $\alpha=0.6$, it can explain the observed radio spectrum slightly steeper than the simplest prediction of $\alpha=0.5$ \citep{rey92}.

Finally we mention the case where an SNR is in a hot wind bubble with a lower density n $\sim 10^{-1-2}{\rm cm}^{-3}$ and a higher temperature T $\sim10^{1-2}$ eV. The growth rate of the resonant instability depends on $n_{\rm p,cold}/n_{\rm p,hot}$ and $\Omega_{\rm c,p}$ \citep{win84}. The growth rate of the Weibel instability depends on $\omega_{\rm p,cold} \propto n_{\rm p,cold}^{1/2}$ and decreases rapidly when the growth rate is comparable to $\Omega_{\rm c,p}$. Hence, for lower upstream density, the growth rate of the resonant instability does not change from that of figure 1, but the growth rate of the Weibel instability becomes smaller than that of figure 1. A higher temperature does not affect much the growth rates because the proton thermal velocity is much smaller than the relative velocity between cold and hot protons. Thus, the mechanism proposed in this letter works as long as the neutral fraction of ISM is as high as $\sim 0.1$.

\acknowledgments
We thank T. Inoue and B. Reville for discussions and useful information about neutral clumps. This work is partly supported by Scientific Research Grants (F.T.: 18542390 and 20540231) by the Ministry of Education, Culture, Sports, Science and Technology of Japan. Y. O. is supported by a Grant-in-Aid for JSPS Research Fellowships for Young Scientists.

\clearpage

\begin{figure}
\plotone{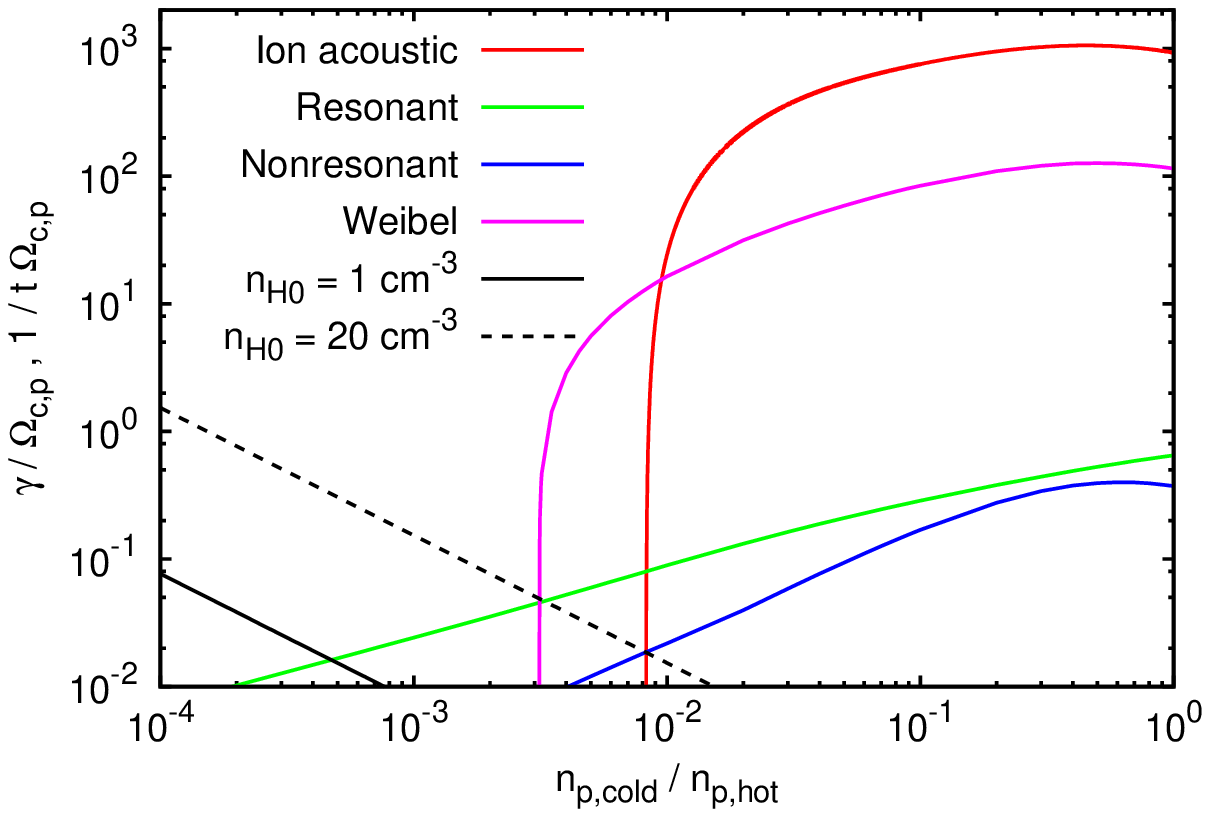} 
\caption{The maximum growth rate of instabilities. The black line and black dash line show ${1}/{t\Omega_{\rm c,p}}$ with $n_{\rm H0}=1{\rm cm}^{-3}$ and $n_{\rm H0}=20{\rm cm}^{-3}$, respectively.
\label{fig1}}
\end{figure}


\begin{thebibliography}{}

\bibitem[Axford et al.(1977)]{axf77}Axford, W. I., Leer, E., \& Skadron, G., 1977, Proc. 15th Int. Cosmic Ray Conf., Plovdiv, 11, 132

\bibitem[Bamba et al.(2005)]{bam05} Bamba, A., Yamazaki R., Yoshida T., Terasawa T., and Koyama, K., 2005, \apj, 621, 793

\bibitem[Blandford \& Ostriker(1978)]{bla78}Blandford, R. D., \& Ostriker, J. P., 1978, \apj, 221, L29

\bibitem[Bell(1978)]{bel78}Bell, A. R., 1978, \mnras, 182, 147

\bibitem[Bell(2004)]{bel04} Bell, A. R., 2004, \mnras, 353, 550

\bibitem[Bell(2005)]{bel05} Bell, A. R., 2005, \mnras, 358, 181

\bibitem[Berezhko et al.(2003)]{ber03} Berezhko, E. G., Ksenofontov, L. T., \& V{\"o}lk, H. J., \aap, 412, L11

\bibitem[Chevalier \& Raymond(1978)]{che78} Chevalier, R. A. \& Raymond, J. C., 1978, \apj, 225, L27

\bibitem[Chevalier et al.(1980)]{che80} Chevalier, R. A., Kirshner, R. P., \& Raymond, J. C., 1980, \apj, 235, 186

\bibitem[Ghavamian et al.(2000)]{gha00}Ghavamian, Raymond, J., Hartigan, P., \& Blair, W. P., 2000, \apj, 535, 266

\bibitem[Ghavamian et al.(2002)]{gha02}Ghavamian, Winkler, P. F., Raymond, J. C., \& Long, K. S., 2002, \apj, 572, 888

\bibitem[Ghavamian et al.(2007)]{gha07}Ghavamian, P., Laming, J. M., \& Rakowski, C. E. 2007, \apj, 654, L69

\bibitem[Giacalone \& Jokipii(2007)]{gia07} Giacalone, J., \& Jokipii, J.R., 2007, \apjl, 663, L41

\bibitem[Heiles \& Troland(2003)]{hei03} Heiles, C., \& Troland, T. H., 2003, \apj, 586, 1067

\bibitem[Heng \& McCray(2007)]{hen07} Heng, K., \& McCray, R., 2007, \apj, 654, 923

\bibitem[Ichimaru (1973)]{ich73}Ichimaru, S.,1973, Basic Principles of Plasma Physics (USA: W.A. Benjamin Inc., Reading, Ma.)

\bibitem[Inoue \& Inutsuka(2008)]{ino08} Inoue, T., \& Inutsuka, S., 2008, \apj, 687, 303

\bibitem[Inoue et al.(2009)]{ino09} Inoue, T., Yamazaki, R. \& Inutsuka, S., 2009, \apj, 695, 825

\bibitem[Kato \& Takabe(2008)]{kat08} Kato, N. T., \& Takabe, H., 2008, \apj, 681, L93  

\bibitem[Koyama et al.(1995)]{koy95} Koyama, K., Petre, R., Gotthelf, E. V., Hwang, U., Matsuura, M., Ozaki, M., \& Holt, S. S. 1995, \nat, 378, 225

\bibitem[Krymsky (1977)]{kry77} Krymsky, G. F., 1977, Doki. Akad. Nauk SSSR, 234, 1306

\bibitem[Lucek \& Bell(2000)]{luc00} Lucek, S. G., \& Bell, A. R. 2000, \mnras, 314, 65

\bibitem[Niemiec et al.(2008)]{nie08} Niemiec, J., Polh, M., \& Nishikawa, K., 2008, \apj, 684, 1174

\bibitem[Ohira \& Takahara (2008)]{ohi08} Ohira, Y., \& Takahara, F., 2008, \apj, 688, 320

\bibitem[Ohira et al.(2009)]{ohi09} Ohira, Y., Reville, B., Kirk, J. \& Takahara, F., 2009, \apj, 698, 445

\bibitem[Reynolds \& Ellison(1992)]{rey92} Raynolds, S. P., \& Ellison, D. C., 1992, \apj, 399, L75

\bibitem[Reville et al.(2006)]{rev06} Reville, B., Kirk, J. G., \& Duffy, P., 2006, Plasma Physics and Controlled Fusion, 48, 1741

\bibitem[Reville et al.(2007)]{rev07} Reville, B., Kirk, J. G., Duffy, P. \& O'Sullivan, S., 2007, \aap, 475, 435

\bibitem[Reville et al.(2008)]{rev08} Reville, B., O'Sullivan, S., Duffy, P. \& Kirk, J. G., 2008, \mnras, 386, 509

\bibitem[Riquelme \& Spitkovsky(2008)]{riq08} Riquelme, M. A. \& Spitkovsky, A., 2008, \apj, 694, 626

\bibitem[Smith et al.(1991)]{smi91}Smith, R. C., Kirshner, R. P., Blair, W. P., \& Winkler, P. F., 1991, \apj, 375, 652

\bibitem[Uchiyama et al.(2007)]{uch07} Uchiyama, Y., Aharonian, F A., Tanaka, T., Takahashi, T., \& Maeda, Y., 2007, \nat, 449, 576

\bibitem[van Adelsberg et al.(2008)]{van08} van Adelsberg, M., Heng, K., McCray, R., \& Raymond, J. C., 2008, \apj, 689, 1089

\bibitem[Vink \& Laming(2003)]{vin03} Vink, J., \& Laming, J. M., 2003, \apj, 584, 758

\bibitem[Winske \& Leroy(1984)]{win84}Winske, D., \& Leroy, M. M., 1984, JGR, 89, 2673

\bibitem[Wu \& Davidson(1972)]{wu72}Wu, S. C., \& Davidson, R. C., 1972, JGR, 77, 5399 

\bibitem[{{Zirakashvili} \& {Ptuskin}(2008)}]{zir08}
{Zirakashvili}, V.~N., \& {Ptuskin}, V.~S. 2008, \apj, 678, 939


\end{thebibliography}
\end{document}